\documentclass[journal]{IEEEtran}
\IEEEoverridecommandlockouts

\usepackage{graphicx}
\usepackage{color}
\usepackage{booktabs}
\usepackage{url}
\usepackage{hyperref}
\usepackage[dvipsnames]{xcolor}

\usepackage{graphicx}

\begin{document}

\title{Open Source Horizontal IoT Platforms: \\A Comparative Study on Functional Requirements}

\author{Ali~Farhat,~\IEEEmembership{Graduate Student Member,~IEEE,}
        Abdelrahman Eldosouky,~\IEEEmembership{Member,~IEEE,}\\
        Jason~Jaskolka,~\IEEEmembership{Senior Member,~IEEE,}
        Mohamed Ibnkahla,~\IEEEmembership{Senior Member,~IEEE,}\\
        and Ashraf~Matrawy,~\IEEEmembership{Senior Member,~IEEE,}
\thanks{This work has been supported in part by the Defence Research and Development Canada’s (DRDC) Canadian Safety and Security Program (CSSP), Project \# CSSP-2018-CP-2344.}
\thanks{A. Farhat, A. Eldosouky, J. Jaskolka, and M. Ibnkahla are with the Department of Systems and Computer Engineering, Carleton University, Ottawa, ON, Canada.}
\thanks{A. Matrawy is with the Department of Systems and Computer Engineering, Carleton University, Ottawa, ON, Canada.}
}

\maketitle
\begin{abstract}
The growth in the deployment of Internet of Things (IoT) devices in various industries required the use of IoT platforms to manage, automate and control devices. This introduced different commercial and open source IoT platforms for developers and researchers to deploy. As a result, selecting one of these platforms for a specific application and use case became a challenge. In this study, a guideline for selecting an open source platform is presented. The process starts by identifying a list of functional requirements that would reflect the requirements of an IoT system in general. This list of requirements is used to compare between four major open source platforms: 1) OM2M (OneM2M standard), 2) IoTivity (OCF standard), LwM2M (OMA SpecWorks LwM2M standard), and 4) FIWARE (FIWARE standard). The purpose of this comparison is to indicate the capability and limitations of the different platforms and how they satisfy each requirement. Afterwards, two examples are presented to demonstrate how this guideline is used to select the most suitable platform for an e-health and a smart city use case. This includes how to define each use case and all the required information that could affect the process of selecting the most suitable platform for the development of the IoT platform.

\end{abstract}


\section{Introduction} \label{sec:introduction}

The Internet of things (IoT) introduced the game-changing concept that different categories of objects including sensors, actuators, computers, people, animals, infrastructure, and buildings can interconnect to provide real-time data~\cite{Carolina2018-ADistributed}. This, in turn, benefited in providing more economical and environmentally friendly solutions to the public while improving quality of life~\cite{Carolina2018-ADistributed}. Due to its numerous applications, the adoption of IoT systems has been increasing over the years where the number of connected IoT devices was 10 billion in 2019 and it is expected to reach 30.9 billion in 2025~\cite{IoTAnalytics:Ref1}. This clearly demonstrates the significance of IoT systems. 
Although, IoT has provided solutions to several problems, it also introduced new problems and challenges. IoT systems suffer from privacy and security issues as a result of the different interactions between its heterogeneous devices with different processing capabilities~\cite{Carolina2018-ADistributed,UPUL2019-Machine, IOT:Ref69}. 
Moreover, the growth rate of the connected devices, and generated and stored data, makes it challenging to manage the security issues in IoT~systems.

One approach to manage heterogeneous IoT devices, and their interactions and data is through the use of a horizontal IoT platform~\cite{IOT:Ref69,IOT:REF56}. Horizontal IoT platforms work by enabling the integration of heterogeneous IoT devices at the data level to establish data interoperability between heterogeneous technologies. This also helps to create a common security management layer that manages all the connected devices. For instance, some IoT devices have simple hardware capabilities, which limit the security and privacy mechanisms that can be implemented on them. Consequently, such devices expose the system and make it vulnerable to attacks as they are considered as part of the system's attack surface~\cite{IOT:Ref68}. Additionally, these connected devices generally use different communication technologies (e.g., Wi-Fi, Bluetooth, etc.), and different communication protocols (e.g., Constrained Application Protocol (CoAP), Message Queuing Telemetry Transport (MQTT), Modbus, etc.). Moreover, the heterogeneity could extend to include devices manufactured by various vendors, which are designed and implemented using different sets of protocols, hardware components, standards, and regulations. In this regard, horizontal IoT platforms establish data interoperability between these heterogeneous devices, regardless of their technologies. As a result, such platforms reduce the potential attack surface of the IoT system as heterogeneity of connected devices is one of the main challenges in IoT.

The main difference between horizontal IoT platforms and traditional vertical platforms is that vertical platforms are built to serve a particular application or to resolve a set of problems within a specific application. In these platforms, the end-user is usually offered with a limited set of options or features specified by the platform. On the other hand, horizontal IoT platforms are general platforms that do not consider a particular IoT application but rather, they provide the tools and features needed to manage devices and data across various IoT applications. Thus, horizontal IoT platforms can be found in different IoT applications such as energy sectors, e-health, transportation, supply chain, etc. These platforms can either be commercial platforms (e.g., Amazon AWS, Google Cloud, Microsoft Azure, etc.) or open-source platforms that provide more flexibility and interoperability in connecting different IoT devices. For instance, in open-source platforms, developers can build their own Application Programming Interfaces (APIs) to connect new device categories, provide extra features, and implement security solutions. In this work, the focus will be on open-source horizontal IoT platforms as they represent an active field of research compared to commercial platforms. However, since there are multiple horizontal IoT platforms that differ in the functionalities they offer and support, selecting a suitable platform for a specific application or project becomes a challenging task.

To this end, the contribution of this work is to: 

\begin{itemize}
    \item Survey and present the major open source horizontal IoT platforms.
    
    \item Present an in-depth set of functional requirements for a horizontal IoT platform. These requirements are used to compare between the presented horizontal IoT platforms and their standards. The presented requirements and comparison shall provide developers and researchers with a guideline for selecting the most suitable open source horizontal IoT platform for their use cases.
    
    \item Demonstrate how to use the presented platform requirements and comparison guideline in selecting the most suitable open source horizontal IoT platform for an e-health use case and a smart city use case. 
\end{itemize}

The remaining sections of this paper are organized as follows: Section~\ref{sec:LiteratureReview} summarizes the related works. Section~\ref{sec:HorIoTPlatform} presents the open source horizontal IoT platforms. The proposed requirements for the horizontal IoT platform are reported in Section ~\ref{sec:req}. The comparison between the presented horizontal IoT platforms based on the introduced requirements is reported in Section~\ref{sec:comparison}. The outcome of the comparison between the horizontal IoT platform is discussed in Section~\ref{sec:discussion}. Afterwards, Section~\ref{sec:usecases} presents two use cases as examples illustrating how the proposed requirements and comparison could be utilized by researchers, developers and practitioners to define the use cases and select the most suitable platform for their application. Finally, Section~\ref{sec:conclusion} concludes the paper.

\section{Related Works} \label{sec:LiteratureReview}

Several researchers have compared different IoT platforms as their need became a necessity. These comparisons considered different evaluation criteria for platforms in terms of their management, and functional requirements~\cite{IOT:REF56, IOT:REF52, IOT:REF57}.

The work in~\cite{IOT:REF56} surveyed and discussed IoT management protocols and frameworks. The work presented the architecture of frameworks and their management aspects. Therefore, it does not reveal the drawbacks and limitations of these frameworks. However, it discussed and compared between the management protocols while discussing the capability and limitation of each protocol. As a result, the comparison of this work is limited when it comes to a developer that needs to understand the drawbacks and limitations of the presented frameworks. A similar comparison is found in~\cite{IOT:REF57} where the comparison presented the reference architectures of different IoT platforms. Consequently, this work provides the details of the comparison between the considered  platforms based on the identified requirements. This highlights each platform's degree of satisfaction for each requirement to guide researchers and developers to identify the capabilities and limitations of these platforms.

The authors in~\cite{IOT:REF52} compared between commercial IoT platforms based on their introduced list of requirements for IoT platforms. Multiple quantitative comparison methods were used the identified functional requirements to conduct the comparison. They used a stackchart, statistical tests (Anova test, error bar test, one-way Anova test, and Tukey’s honest significant difference test), multi-criteria decision making based on analytical hierarchical process, and clustering analysis. Finally, they suggested an approach to select the suitable IoT platform. However, the presented quantitative comparisons consider the scenario where all the requirements are treated with equal significance. However, this is not the case in a real-life scenario since the requirements might have different priorities set by the developers. Therefore, if different weights are set for different requirements, the selected platform could be different based on the comparison. This could be solved by stating the details of the comparison and how each platform meets a requirement. Consequently, developers could follow the same approach followed by the authors while considering the details provided in the comparison and utilize their own weights. The result would help the developers select the option that would require the least amount of additional work given that it will be able to satisfy the end goal.

On the other hand, there are several processes in the literature that introduces IoT system requirements or defines a complete use case. The work in~\cite{IOT:REF58} presented IoTReq, a Unified Modeling Language (UML)-based process to determine the requirements and specifications of an IoT system for a specific use case. It starts by modelling the domain using UML based on a service-oriented method. Afterwards, the goals of the proposed IoT system are identified to extend the created UML model to produce the functional requirements of the IoT system. Another group of researchers in~\cite{IOT:REF53} developed a requirements engineering process that helps the developers define their use case and its requirements in a systematic approach. Their presented process is a customized process to focus on IoT systems from the original processes of ISO IEC/IEEE 12207:2017. The aim of their approach is to ensure that the defined requirements achieve the expected needs of the IoT platform. The presented process consists of three main sub-processes: 1) project scope definition, 2) IoT system definition, and 3) IoT system requirements definition. Each sub-process consists of several cohesive activities and tasks that aims to achieve the expected outcome. After identifying the scope of the project in the first sub-process, the IoT system definition and IoT system requirements definition sub-processes are iterative sub-processes. In other words, they are executed continuously until the final outcome meets the expectation. This process could be used by the developers to ensure that their defined horizontal IoT platform requirements are leading them to a platform that satisfies their need. 

\section{Horizontal IoT Platforms} \label{sec:HorIoTPlatform}
In this section, the considered horizontal IoT platforms are introduced after discussing the advantages of open source and standard-based platforms.

\subsection{General Characteristics}
\subsubsection{Open Source}

The need for an open source solution in IoT development is due to three main reasons~\cite{OCF:Ref2}: 1) Time saving and Customization, 2) Expertise, and 3) Scalability and Reliability. Starting the development through a mature open source implementation means that the developers are reusing available solutions instead of “reinventing the wheel” for every product. In addition, product developers may not have the low-level software expertise required to develop every single aspect of the product. Therefore, the presence of open source IoT infrastructure allows the developers of a certain product to benefit from a robust and reliable underlying communications stack and application codes and focus on enhancing their applications. Subsequently, such open source components reduce what is known as “time to market” and develop the product in a short period of time. Another reason to go after open source is the expertise where the developers can reuse mature components that will help avoid any common pitfalls in the early development. Such components have been extensively reviewed by experts in the field where improvements are identified and addressed by a large community of contributors. For example, industry standards such as Open Connectivity Foundation (OCF) and OneM2M are established by experts to create an open source baseline that is interoperable and offered as a proven solution. Additionally, one of the main aims of development in IoT is scalability and reliability. The fact that IoT devices have specific needs such as lightweight messaging for sensor networks, and low-power high range of applications should be considered when developing the device. Open source IoT solutions are developed by experts that are aware of these needs and consider them in addition to the vertical and horizontal components of IoT. Moreover, the open source implementations help demonstrate that the solution “works” in advance and before adoption~\cite{OCF:Ref2}.  

\subsubsection{Standard Compliance}

The need for a horizontal platform that satisfies a standard is critical since industry organizations establish standards that ensure interoperability and offer the open source software to provide the proven basis in the IoT market. In the IoT industry, security is considered of utmost importance and the solution should address security proactively and at its early design stage instead of patching up its vulnerabilities at later stages of development. Therefore, experts in the area of security architecture and implementation dedicate a substantial amount of time on creating the robust baseline to ensure that the open source component is not the “weak link” of the IoT ecosystem. Moreover, such foundations enable IoT interoperability through developing industry standards allowing IoT devices to communicate effectively and securely. Consequently, developers will be able to easily design and scale their devices. 
 
\subsection{Horizontal IoT Platforms Reference Implementations}\label{sec:refImp}
There are several horizontal platforms in IoT, however, not all platforms are open source or implement a standard. In this section, open source platforms that implement a specification standard for Machine-to-Machine (M2M) communication are presented. 

\subsubsection{Eclipse OM2M}

Eclipse OM2M is an open source implementation of the oneM2M and SmartM2M standards. The project is initiated by the Laboratory for Analysis and Architecture of Systems - Centre National de la Recherche Scientifique (LAAS-CNRS) to provide a horizontal M2M service platform. It aims to provide the ability to develop services independently of the underlying network while facilitating the deployment of vertical applications and heterogeneous devices~\cite{TEF:Ref3}. It is worth noting that the considered OM2M platform in the comparison is the platform that is based on the OneM2M standard since it is the most recent and in active development. The development of the OM2M platform that is based on the SmartM2M standard is no longer active. 

To enable such development, the OM2M platform follows a RESTful approach with open interfaces. Its modular architecture runs on the top of an Open Service Gateway Initiative (OSGi) layer which makes it highly extensible via plugins. It supports several protocol bindings such as HyperText Transfer Protocol (HTTP), MQTT, and CoAP. Additionally, it provides various interworking proxies that enable seamless communication. It provides a horizontal Common Services Entity (CSE) which is deployed in the server, gateway, and devices. The CSE provides application enablement, security, triggering, notification, persistency, device interworking, device management, etc.~\cite{TEF:Ref3}. The modular architecture of the OM2M platform makes the platform highly extensible via plugins. This allows the developers to create their own contributions to the platform to add new features.

\subsubsection{IoTivity}

IoTivity is an open source framework that implements the OCF standard to perform easy and secure IoT device-to-device communication. IoTivity includes a reference implementation of a client software known as the Onboarding Tool and Generic Client (OTGC), an application for controlling the IoT device and it is available on Linux and Android. Additionally, it includes a reference implementation of a server software, which runs on the IoT device itself. Moreover, it provides the tools that generate the code for the server through a JavaScript Object Notation (JSON) file, which is created to describe the device’s capability. It is worth noting that IoTivity v2.1.2 was tested for compliance with the OCF 2.1.2 specifications and both are considered in the comparison~\cite{IOTV:Ref4}. 

\subsubsection{Lightweight M2M (LwM2M)}

LwM2M is a client-server-based device management protocol developed by OMA SpecWorks for sensor networks. It aims to satisfy the demand of the market by establishing a common standard that manages resource constrained devices to help realizing the potential of IoT. It is designed to manage devices and related services remotely~\cite{OMA:Ref5}.  

There are several opensource implementations of the OMA SpecWorks LwM2M: 
\begin{itemize}
    \item \textbf{OMA LwM2M DevKit}: The OMA LwM2M DevKit is a Mozilla Firefox web browser add-on. It adds the support of the LwM2M protocol and enables the manual interaction with the LwM2M server~\cite{OMA:Ref6}.
    
    \item \textbf{Eclipse Leshan}: Eclipse Leshan is a Java implementation of LwM2M that provides libraries to help developers develop their own LwM2M server and client~\cite{EF:Ref7}.
    
    \item \textbf{Eclipse Wakaama}: Eclipse Wakaama is a collection of files written in C and intended to be built in an application. It provides APIs for a server application to communicate with registered lightweight clients, where Wakaama on the client applications checks the received commands for syntax and access rights and dispatch them to the relevant objects. It is worth noting that it is mono-threaded~\cite{EF:Ref8}. 
    
    \item \textbf{Anjay}:  Anjay is a set of tools that enables device vendors to implement a LwM2M client on their hardware or develop their customized LwM2M client for testing purposes~\cite{AV:Ref9}. The open source version of Anjay implements the LwM2M specification version 1.0.2, where the commercial version implements version 1.1 of the specification~\cite{AV:Ref10}.
\end{itemize}

\subsubsection{FIWARE}
FIWARE is an open source initiative that defines a universal set of standards for context data management facilitating the development of smart solutions in various domains~\cite{FI:Ref11}. The FIWARE platform consists of multiple sets of general-purpose functions known as general enablers that offer functions independent from any usage area~\cite{FI:Ref12}. These generic enablers provide the developers with open interfaces through APIs and support interoperability with other generic enablers. To build any smart solution while utilizing FIWARE, it will be built around the FIWARE context broker generic enabler, the only mandatory component of the FIWARE platform. However, there are components in the complementary suite of the FIWARE platform that can be used to extend the solution and help in:
\begin{itemize}
    \item Interfacing with other system components including IoT, robots, and third-party systems to update the context information, and perform corresponding actions.
    \item Context data, API management, and publication and monetization to support usage control.
    \item Process, analyze, and visualize context information to assist applications and users in smart decision making and meet the expected smart behavior. 
\end{itemize}

Similar to the OM2M platform, the FIWARE platform provides the developers with the ability to add their own contribution. In this case, the platform would consist of the context broker generic enabler, the platform's only mandatory component that manages context information, and additional complementary generic enablers. These additional complementary generic enablers can be added or combined with other third-platform components to design the hybrid platform that satisfies the requirements of the use case. This allows the developers to label the platform as a “Powered by FIWARE” platform. The advantage of using FIWARE compared to other alternatives is the availability of ready made generic enablers that process, analyze, and visualize collected data. In addition, these enablers stream FIWARE's database to other specialized data processing platforms such as Amazon~AWS and Spark. 

It is worth noting that the source of the reported information for the FIWARE platform in the comparison is the official website of the platform as it is the source of documentation.

\section{Horizontal IoT Platform functional requirements} \label{sec:req}
The identification of requirements (both functional and non-functional requirements) is a critical step in the development of a platform for an IoT application. 
In particular, the functional requirements describe what the platform must do in terms of its behavior and the features that it will provide to the IoT application. This work focuses on presenting the functional requirements essential to most IoT applications that depend on horizontal IoT platforms. The presented list of requirements consists of 16 functional requirements that are defined in consistency with other works from the literature~\cite{IOT:REF52,IOT:Ref67} while considering the authors' perspective. This list can be used as a reference to compare different horizontal IoT platforms and to select a suitable horizontal platform. In addition, developers and researchers can add to or modify the presented list of requirements according to their specific application or use case.

In this section, the list of functional requirements is introduced. Later, this work presents two examples of how these lists are used to select the most suitable horizontal IoT platform for two different use cases in Section \ref{sec:usecases}. This includes categorizing the functional requirements following the MoSCoW prioritization criteria~\cite{IoT:moscow} to identify the most important requirements for each use case to help select the most suitable platform. The MoSCoW prioritization criteria is separated into four categories of priority: ``Must Have", ``should have", ``Could Have", and ``Will Not Have". The ``Must Have" category refers to the requirements that are mandatory and non-negotiable for the platform. The non-delivery of the ``Must Have" requirements renders the platform improper for the considered application. The ``Should Have" category represents important requirements that may be satisfied through a workaround. Ignoring these requirements might affect the end product, however it may still viable. The ``Could Have" category refers to the desirable and wanted requirements but they are not necessary. They may improve the user experience and satisfaction level of the end product. Finally, the ``Will Not Have" category of requirements represents the requirements not appropriate for this time frame of development.

The requirements below are listed by their identification number (ID) and the description of each requirement:
\begin{itemize}
    \item \textbf{FNR-001}: The platform shall connect an IoT device seamlessly and communicate data (continuously) regardless of the networking technologies (Bluetooth, Bluetooth Low Energy, Wi-Fi, etc.). 
    
    \item \textbf{FNR-002}: The platform shall utilize two-way communication if it is supported by the communication protocols used by the connected IoT device. 
    
    \item \textbf{FNR-003}: The platform shall identify the supported communication technologies and\textbackslash{or} protocols by the IoT device during device registration. 
    
    \item \textbf{FNR-004}: The platform shall protect the confidentiality, integrity, availability, and privacy of data exchanged in transit, at rest, and in processing. 
    
    \item \textbf{FNR-005}: The platform shall be able to perform data processing on the collected data from IoT devices (i.e., data conversion, reformatting, and data analysis). 
    
    \item \textbf{FNR-006}: The platform shall be configurable where possible, allowing features, including security features, to be added, modified, and deactivated.  
    
    \item \textbf{FNR-007}: The platform shall authenticate devices every time they initiate a connection to the platform and before they communicate data. 
    
    \item \textbf{FNR-008}: Users shall be authenticated before accessing the platform every time the user is trying to log in to the platform. 
    
    \item \textbf{FNR-009}: The platform shall log every event that occurs in the platform for auditing purposes. 
    
    \item \textbf{FNR-010}: The platform shall be able to handle traffic with different priorities and characteristics. 
    
    \item \textbf{FNR-011}: The platform shall signal a warning alarm or alert to indicate initialization failure. The Initialization process is the setup process of the device in the platform. This may involve physically attaching or placing the device, registering the device, setting up the communications between the device and the platform, setting up the communications capabilities of the device in the platform to ensure the use of the most secure communications.
    
    \item \textbf{FNR-012}: The platform shall implement an access control mechanism to protect the resources. 
    
    \item \textbf{FNR-013}: The platform shall be able to provide time stamps to the different entities of the platform. 
    
    \item \textbf{FNR-014}: The platform shall be able to synchronize time within the system and between its different entities. 
    
    \item \textbf{FNR-015}: The platform should support location tracking for remote devices if it is supported by the device.
    
    \item \textbf{FNR-016}: The platform should support device maintenance such as software updates. 
\end{itemize}

\section{Comparing Horizontal IoT Platforms Reference Implementations} \label{sec:comparison}

In this section, a comparison between the presented platforms in Section \ref{sec:refImp} based on how they meet the functional requirements identified in Section \ref{sec:req} is conducted. In the comparison, the specific reference implementation for each of the of standards presented in Section \ref{sec:refImp} are considered. For instance, \emph{OM2M} is considered as a reference implementation for the OneM2M standard, \emph{IoTivity} for OCF, \emph{Lightweight M2M} for the OMA SpecWorks Lightweight M2M, and \emph{FIWARE} as it is both a reference implementation and a standard. The comparison will summarize how each platform satisfies a requirement. 

\begin{itemize}
    \item \textbf{FNR-001}: \begin{itemize}
    \item All the reference implementations were found to be able to connect the devices seamlessly. However, it is worth noting that the current implementation of the OM2M platform does not support streamed communication of data (i.e., video and data feeds)~\cite{OneM2M:Ref18}.
    \end{itemize}
    
    \item \textbf{FNR-002}: \begin{itemize}
        \item \textbf{OM2M}: The platform supports the CoAP and MQTT protocols. The platform support mechanisms to confirm delivery of a message and to detect failure of message delivery within a given time interval~\cite{OneM2M:Ref18}. 
        
        \item \textbf{IoTivity}: The platform supports the CoAP and MQTT protocols~\cite{OCF:Ref19}.  
        
        \item \textbf{Lightweight M2M}: The platform supports the CoAP protocol~\cite{OMA:Ref5}. 
        
        \item \textbf{FIWARE}: The platform supports the CoAP and MQTT protocols~\cite{FI:Ref21}. 
    \end{itemize}
    
     \item \textbf{FNR-003}: \begin{itemize} 
        \item \textbf{OM2M}: The platform provides a RESTful API to create and manage M2M resources. This includes several procedures including resources discovery~\cite{OM2M:Ref22}. The resource discovery is accomplished based on their search strings using the discovery resource~\cite{OM2M:Ref23}. 
     
        \item \textbf{IoTivity}: The platform’s resource discovery is part of the base layer and it consists of two main functionalities~\cite{IOTV:Ref24}:  resource registration and resource finding. The discovery is a client-server operation, where the device acting as a server registers its public resources with the stack. Then, a device acting as a client will be able to discover the resources through sending a multicast or a unicast discovery request~\cite{IOTV:Ref24}. 
        
        \item \textbf{Lightweight M2M}: The requirement for the resource discovery is derived from the OneM2M requirements~\cite{OMA:Ref25}. The M2M interface enabler shall be capable of allowing the M2M service layer to discover devices that are managed by the device management server. However, there is no evidence to ensure its implementation. More investigation is needed. 
        
        \item \textbf{FIWARE}: The FIWARE platform has several available GE that interface IoT devices, robots and third-party systems with the platform. For example, the IDAS GE is an implementation of the backend device management. It includes several IoT agents that interface IoT devices supporting the most common and widely used IoT protocols (e.g., LwM2M, JSON, Ultralight, etc.). These agents support several IoT protocols with modular architecture. Their purpose is to connect IoT devices/gateways to FIWARE-based ecosystems through translating IoT-specific protocols into the NGSI context information protocol that is used as the standard data exchange model in the FIWARE platform. All devices must be provisioned before being used by the platform. To do that, a pre-provisioning request is sent which includes the information about the device including the utilized communication protocol by the device~\cite{FI:Ref33,FI:Ref59,FI:Ref60}. On the other hand, FIWARE-ready IoT devices are the devices that natively support the NGSI API and demonstrate the seamless interoperability with the FIWARE platform~\cite{FI:Ref61, FI:Ref62}.

     \end{itemize}
     
     \item \textbf{FNR-004}: \begin{itemize} 
     
        \item \textbf{OM2M}: The platform’s security function layer of the platform contains security functions that can be classified into six categories~\cite{OneM2M:Ref26}: identification, authentication, authorization, security association, sensitive data handling and security administration. The platform’s security service layer provides the following services as part of sensitive data handling~\cite{OneM2M:Ref26}: sensitive functions protections, secure storage, and data error detection. The Transport Layer Security Pre-Shared Key cipher suites (TLS-PSK) protocol is used to secure M2M communications based on pre-shared keys~\cite{OM2M:Ref27}. 
        
        \item \textbf{IoTivity}: The OCF standard emphasizes that OCF devices must support (Datagram) Transport Layer Security ((D)TLS) version 1.2 or greater~\cite{OCF:Ref28}. Sensitive data stored in volatile and non-volatile memory shall be encrypted using acceptable algorithms to prevent access by unauthorized parties~\cite{OCF:Ref28}.
        
        \item \textbf{Lightweight M2M}: The security common functions must be able to provide data confidentiality and integrity protection for any transmitted information between any resources in the network~\cite{OMA:Ref29}.  For example, the platform supports Datagram Transport Layer Security (DTLS) protocol to ensure authentication confidentiality, and integrity protection for data in transit~\cite{OMA:Ref29}. 
        
        \item \textbf{FIWARE}: It is not clear if there is any data protection in the original FIWARE platform. However, the work in \cite{FI:Ref63} shows that it is possible to add the required security mechanisms to protect the data of the platform.The work implemented DTLS version 1.2 as part of the IoT agents used for the LwM2M and CoAP protocols.  
     
     \end{itemize}
     
     \item \textbf{FNR-005}: \begin{itemize} 
        
        \item \textbf{OM2M}: Data aggregation, format conversion and storage for analysis is provided by the platform’s data management and repository~\cite{OneM2M:Ref30}. 
        
        \item \textbf{IoTivity}: The platform uses Concise Binary Object Representation (CBOR) to marshal a database before storing or sending it. This functionality is available in the security resource manager~\cite{IOTV:Ref31}. It is unclear if the platform supports any data processing. 
        
        \item \textbf{Lightweight M2M}: The platform supports several data formats including JSON, plain text, and binary~\cite{OMA:Ref32}.  It is unclear if the platform supports any data processing. 
        
        \item \textbf{FIWARE}: The platform is capable of processing, analyzing, and visualizing information to provide the expected smart behavior of applications~\cite{FI:Ref33}. However, it is not clear what are the limitations of such processing. The platform provides several Enablers as part of the core context management~\cite{FI:Ref33}: Cygnus generic enabler stream data into popular databases such as MongoDB, and Amazon AWS. Cosmos enabler perform simple Big data analysis with Spark and Flink big data platforms.  The platform provides several enablers to make it easier to interface the platform. The IDAS generic enabler offers a wide range of IoT Agents that makes it easier to interface with devices using the most widely used IoT protocols such as~\cite{FI:Ref33}: LwM2M over CoAP, JSON, and  UltraLight over HTTP/MQTT.
        
     \end{itemize}
     
    \item \textbf{FNR-006}: \begin{itemize} 
    
        \item \textbf{OM2M}: The platform is proposed with a modular architecture running on top of an OSGi layer and it is highly extensible via plugins. This is demonstrated by the work in~\cite{Ref34} where the authors have proposed a policy enforcement layer to deal with security and privacy of a specific application requirements to enhance the robustness of the platform. They proposed a new policy enforcement plugin that managed the access to the resources of the platform and handle any violation attempts of the policies.
        
        \item \textbf{IoTivity}: It is not clear if it is possible to add/ modify/ deactivate features of the platform. 
        
        \item \textbf{Lightweight M2M}: It is not clear if it is possible to add/modify/deactivate features of the platform. 
        
        \item \textbf{FIWARE}: The platform is designed in a way that the only component in the platform that should be used and cannot be changed is the Context Broker. Any other components can be modified and altered and hence designed accordingly~\cite{FI:Ref33}. 
    
    \end{itemize}
    
    \item \textbf{FNR-007}: \begin{itemize}
    
        \item \textbf{OM2M}: The platform performs authentication after validating the identity supplied in the identification step where it is associated with a trustworthy credential. How to perform an authentication process depends on the used mutual authentication mechanism~\cite{OneM2M:Ref26}. For example: A digital signature is required for certificate-based authentication mechanism. A Message Integrity Code (MIC) is required for symmetric key based authentication mechanism.
        
        \item \textbf{IoTivity}: The platform’s server shall authenticate a client (client refers to a device in the role of a client) when it requests access to a restricted resource on a server~\cite{OCF:Ref28}. Additionally, clients shall authenticate servers while requesting access. The authentication can be accomplished using: Symmetric key credentials, Raw asymmetric key credentials, and Certificates.
        
        \item \textbf{Lightweight M2M}: The OMA security gateway provides server security services to the resources connected to it including authentication~\cite{OMA:Ref35}. For mutual authentication, there are three cases~\cite{OMA:Ref35}: A shared key is used together with PSK-TLS, the server is authenticated via a server certificate and the security agent via HTTP Digest, and the server and the client both use certificates for authentication. 
        
        \item \textbf{FIWARE}: The platform authenticates devices using the Open Authorization 2 (OAuth2)-based authentication~\cite{FI:Ref33}. 
        
    \end{itemize}

    \item \textbf{FNR-008}: \begin{itemize}
    
        \item \textbf{OM2M}: The platform authenticates users using PIN numbers. PINs are used to identify the owner of an Asymmetric Secure Element (ASE) and to protect its data where the data is accessed upon successful verification of the PIN. The activation PIN is needed only once during the operational phase~\cite{OneM2M:Ref26}. 
        
        \item \textbf{IoTivity}: The platform’s server examines the resource requested by the client before processing the request by searching the Access Control Element (ACE) to find one or more entries that matches the client and the requested resource. If it is found, each entry is evaluated independently~\cite{OCF:Ref28}. 
        \item \textbf{Lightweight M2M}: The OMA Security Gateway provide server security services to the resources connected to it including authentication~\cite{OMA:Ref35}.  For mutual authentication, there are three cases~\cite{OMA:Ref35}: A shared key is used together with TLS-PSK,  the server is authenticated via a server certificate and the security agent via HTTP Digest, and the server and the client both use certificates for authentication. 
        
        \item \textbf{FIWARE}: The platform authenticates users using the OAuth2-based authentication~\cite{FI:Ref33}. 
    
    \end{itemize}

    \item \textbf{FNR-009}: \begin{itemize}
    
        \item \textbf{OM2M}: The platform uses the “eventlog” resource to log all the events related to a device~\cite{OneM2M:Ref30}. 
        
        \item \textbf{IoTivity}: The platform can generate various kinds of auditable events where they can be used for log analysis in a real-time to help understand the system condition and backtrack any issue that have occurred in the system. However, since the storage capacity of IoT devices is typically very limited, a data structure such as a ring buffer is often used~\cite{OCF:Ref28}.  For every auditable event entry, the timestamp property consists of a full-date and partial-time format. For every new auditable event, its timestamp should have a later time compared to the latest previously logged auditable event~\cite{OCF:Ref28}. 
        
        \item \textbf{Lightweight M2M}: The log information is transferred from the device to the LwM2M server based on a request from the server~\cite{OMA:Ref36,OMA:Ref37}.
        
        \item \textbf{FIWARE}: The Orion context broker of the FIWARE platform provides a log file. The following different error messages are printed to the log file based on the selected option \cite{FI:Ref64}:
        
        \begin{itemize}
            \item NONE: No log at all.
            \item FATAL: Only FATAL ERROR messages are logged.
            \item ERROR: Only error messages are logged.
            \item WARN (default): Warning and error messages are logged.
            \item INFO: Information, warning and error messages are logged.
            \item DEBUG: Debug, information, warning and error messages are logged.

        \end{itemize}

        The log file format is designed to be processed by tools such as Splunk \cite{FI:splunk} or Fluentd \cite{FI:fluentd} where each line is composed by several key value fields separated by the pipe character (\textbar) \cite{FI:Ref64}.
        
        Each log message includes timestamp, log message level, correlator ID, transaction ID, source Internet Protocol (IP) address, associated service to the transaction, component, the source code function generating the log message, and the content of the log message \cite{FI:Ref64}.

    \end{itemize}
    
    \item \textbf{FNR-010}: \begin{itemize}
        
        \item \textbf{OM2M}: The platform prioritizes messages when it is forwarded~\cite{OneM2M:Ref38}. A message is stored in the forwarding buffer with its corresponding storage priority. The value for the storage priority is from 1 to 10. The higher the value, the higher the priority. If the memory is not enough in the forward buffer to store a higher priority message, purge the messages with lower priority~\cite{OneM2M:Ref38}. 
        
        \item \textbf{IoTivity}: The platform prioritizes endpoints. In case there are several endpoints in the system, a value denoted by “pri” indicates the priority among them. The value is given a positive number and the smaller it is, the higher the priority. The default value is 1 which is equivalent to not having a priority~\cite{OCF:Ref19}. 
        
        \item \textbf{Lightweight M2M}: The prioritization is based on the quality of service~\cite{OMA:Ref39}. 
        
        \item \textbf{FIWARE}: It is unclear if the platform handle traffic with different priorities. 
        
    \end{itemize}
    
       \item \textbf{FNR-011}: \begin{itemize}
        
        \item \textbf{OM2M}: The platform responds with an unsuccessful response with a response status code to the originator when a subscription to a specific resource fail~\cite{OneM2M:Ref38}.  
        
        \item \textbf{IoTivity}: The platform provides the current status of a device and its corresponding last error code. The last error code indicates for the failure reason~\cite{IOTV:Ref40}.  For example, if the code "0" means there is no error, the code "1" indicates that a given Service Set IDentifier (SSID) is not found, and the code "2" means that the Wi-Fi’s password is wrong.
        
        \item \textbf{Lightweight M2M}: The platform has a device management client-server protocol. It provides a formal interface that allows the server to send device management commands to clients and clients may return status and alerts to servers~\cite{OMA:Ref41}. 
        
        \item \textbf{FIWARE}: The platform provides an error response with a JSON payload that includes a required textual description of the error, and an optional additional description about the error~\cite{FIWARE:Ref42}. Additionally, the platform provides warnings in the log file. For more details on the log file, refer to the information of the FIWARE platform in the FNR-009 requirement. 
        
    \end{itemize}
    
    \item \textbf{FNR-012}: \begin{itemize}
        
        \item \textbf{OM2M}: The platform’s access control approach conforms to the concept of Attribute-Based Access Control (ABAC)~\cite{OneM2M:Ref26}.  The Access Control Polices (ACP) are used by the CSE to control access to the resources. The resources are linked to the ACPs where these ACPs are shared among resources. They contain the privileges which define~\cite{OneM2M:Ref26}:  who can access the resource,  what is the allowed operation (create/retrieve/delete/update, etc.), and under which contextual constraint (access time, access location and IP address, etc.).
        
        \item \textbf{IoTivity}: The platform’s access control mechanism may follow~\cite{OCF:Ref28}: Subject-Based Access Control (SBAC), Role-Based Access Control (RBAC), and  Wildcard-Based Access Control. For every resource instance in the platform, it is required to have an associated access control policy. Hence, for every device acting as a resource server, there is an Access Control List (ACL). In case the access control is using subject-based, it is needed to have an ACE for each subject (identity of a client) that needs to access a subject-based controlled resource.  It may be possible and subject to default permissions defined for the resource for unknown or anonymous (unauthenticated) subject. More details of the format for the ACL can be found in~\cite{OCF:Ref28}. 
        
        \item \textbf{Lightweight M2M}: The platform does mention Access Control, but it is not clear what is the mechanism More investigation is needed~\cite{OMA:Ref43}. 
        
        \item \textbf{FIWARE}: The platform can implement the following two enablers~\cite{FI:Ref33}: AuthZForce PDP/PAP and  Wilma proxy. Both enablers implement access control schema based on the eXtensible Access Control Markup Language (XACML) standard (ABAC)~\cite{FI:Ref33}.
        
    \end{itemize}

      \item \textbf{FNR-013}: \begin{itemize}
        
        \item \textbf{OM2M}: The platform’s data management and repository provide the ability to store contextual information such as timestamps, location, datatypes, etc.~\cite{OneM2M:Ref30}. However, there is no clear evidence that the platform provides time stamps. 
        
        \item \textbf{IoTivity}: The platform utilize timestamps and it consists of a full-date and partial-time format~\cite{OCF:Ref19}.
        
        \item \textbf{Lightweight M2M}: The server writes the value of time at the device. Then, it is the responsibility of the device to increase the value as every second elapses~\cite{OMA:Ref43}. However, there is no clear explanation if this is used to generate time stamps. 
        
        \item \textbf{FIWARE}: The timestamp is provided to the platform as a metadata attribute to represent the time the attribute value was measured. This is part of the main elements in the NGSI data model \cite{FIWARE:Ref42}.
    \end{itemize}

    \item \textbf{FNR-014}: \begin{itemize}
        
        \item \textbf{OM2M}: According to the requirements documentation of the OneM2M standards~\cite{OneM2M:Ref18}, the platform should support time synchronization with an external clock source, and the M2M devices and gateways should support time synchronization within the system. However, it is worth noting that the current implementation of the OM2M platform does not support these time synchronization features~\cite{OneM2M:Ref18}.
        
        \item \textbf{IoTivity}: It is not clear if the platform has any capability to synchronize the timing within the system. 
        
        \item \textbf{Lightweight M2M}: The server writes the value of time at the device. Then, it is the responsibility of the device to increase the value as every second elapses~\cite{OMA:Ref43}. 
        
        \item \textbf{FIWARE}: It is unclear if the platform synchronizes time within the entities of the system. However, the platform provides the current time as a timestamp to the attributes when the attribute values are updated without a timestamp \cite{FI:Ref65}. Additionally, the platform provides the ability to set the timezone of a context entity (e.g. IoT device) as one of the context attributes of the NGSI data model \cite{FI:Ref66}. 
    \end{itemize}
    
    \item \textbf{FNR-015}: \begin{itemize}
        
        \item \textbf{OM2M}: The platform provides a geolocation module that can be associated to set the geolocation of a device This includes mandatory and optional fields including latitude and longitude coordinates, altitude, etc.~\cite{OneM2M:Ref46}. 
        
        \item \textbf{IoTivity}: The platform provides a property that is associated with devices and resources connected to the platform to track the geolocation. This includes mandatory and optional fields including latitude and longitude coordinates, altitude, accuracy, etc.~\cite{OCF:Ref47}. 
        
        \item \textbf{Lightweight M2M}: he platform supports location tracking through the location information part of the \emph{geopriv} object which can be associated with a device. The location could be a geographical location or a civic address~\cite{OMA:Ref48}. 
        
        \item \textbf{FIWARE}: The platform supports location tracking through two formats of geospatial properties~\cite{FIWARE:Ref42}: Simple location Format, a very lightweight format to use by developers and users to add to their entities and GeoJSON which is a data interchange format based on JSON. Developed to provide greater flexibility to represent single point location of more complex geo-spatial shapes such as geometric shapes. 
    \end{itemize}
    
    \item \textbf{FNR-016}: \begin{itemize}
        
        \item \textbf{OM2M}: The platform supports device maintenance such as software updates~\cite{OneM2M:Ref18}.  The platform has the option of using CoAP Block to transfer large payloads that may be used for firmware and software updates~\cite{OneM2M:Ref44}. The details regarding device maintenance and software updates are not clear.
        
        \item \textbf{IoTivity}: The platform supports a secure software update methodology for devices. The update check and install process could be accomplished via the device it-self or a sufficiently privileged client~\cite{OCF:Ref28}. The state of all resources implemented in the device should be the same as after boot, meaning that the software update is not resetting user data and retaining a correct state~\cite{OCF:Ref28}. 
        
        \item \textbf{Lightweight M2M}: The platform has an update management object that provides an interface between the mobile devices (client) and the server to support firmware updates. For each single firmware update, the parameters form a firmware management object. Every management object may contain or refer to one update package~\cite{OMA:Ref45}.  The platform has a client-initiated firmware update option that can be used to update devices and servers. If a client initiated this option, an information message will be sent to the server to verify if an update is needed~\cite{OMA:Ref45}. 
        \pagebreak
        \item \textbf{FIWARE}: It is unclear if the platform supports device maintenance such as software updates.
    \end{itemize}

\end{itemize}

\section{Reference Implementations Discussion} \label{sec:discussion}

After discussing the requirements and the capability of each platform, the findings of the comparison is summarized in Table \ref{Table:tab1}, where the terms correspond as follow: 
\begin{itemize}
    \item \textit{\color{NavyBlue}{Satisfied}}: The platform satisfies the requirement. 
    \item \textit{\color{Orange}{Partially}}: The platform partially satisfies the requirement. 
    \item \textit{\color{Red}Not Satisfied}: The platform does not satisfy the requirement. 
    \item \textit{Unknown}: It is not clear if the platform satisfies the requirement.
\end{itemize}

\begin{table}[!ht]
\centering
\caption{A summary of the horizontal IoT platforms comparison}\label{Table:tab1}
\resizebox{\linewidth}{!}{%
\begin{tabular}{c||c|c|c|c}
\hline
Req. ID        & OM2M                & IoTivity                 & Lightweight M2M      & FIWARE              \\ \hline\hline
FNR-001 & \textbf{\color{NavyBlue}{Satisfied}}           & \textbf{\color{NavyBlue}{Satisfied}}          & \textbf{\color{NavyBlue}{Satisfied}}           & \textbf{\color{NavyBlue}{Satisfied}}           \\ \hline
FNR-002 & \textbf{\color{NavyBlue}{Satisfied}}           & \textbf{\color{NavyBlue}{Satisfied}}          & \textbf{\color{NavyBlue}{Satisfied}}          & \textbf{\color{NavyBlue}{Satisfied}}           \\ \hline
FNR-003 & \textbf{\color{NavyBlue}{Satisfied}}          & \textbf{\color{NavyBlue}{Satisfied}}          & \textbf{\color{NavyBlue}{Satisfied}}           & \textbf{\color{NavyBlue}{Satisfied}}             \\ \hline
FNR-004 & \textbf{\color{Orange}{Partially}}          & \textbf{\color{Orange}{Partially}}             & \textbf{\color{Orange}{Partially}}            & \textbf{\color{Orange}{Partially}}             \\ \hline
FNR-005 & \textbf{\color{Orange}{Partially}}           & \textbf{\color{Orange}{Partially}}             & \textbf{\color{Orange}{Partially}}            & \textbf{\color{NavyBlue}{Satisfied}} \\ \hline
FNR-006 & \textbf{\color{Orange}{Partially}}           & Unknown             & Unknown              & \textbf{\color{NavyBlue}{Satisfied}}           \\ \hline
FNR-007 & \textbf{\color{NavyBlue}{Satisfied}}           & \textbf{\color{NavyBlue}{Satisfied}}          & \textbf{\color{NavyBlue}{Satisfied}}          & \textbf{\color{NavyBlue}{Satisfied}}           \\ \hline
FNR-008 & \textbf{\color{NavyBlue}{Satisfied}}           & \textbf{\color{NavyBlue}{Satisfied}}          & \textbf{\color{NavyBlue}{Satisfied}}           & \textbf{\color{NavyBlue}{Satisfied}}          \\ \hline
FNR-009 & \textbf{\color{NavyBlue}{Satisfied}}           & \textbf{\color{NavyBlue}{Satisfied}}           & \textbf{\color{NavyBlue}{Satisfied}}         & \textbf{\color{NavyBlue}{Satisfied}}              \\ \hline
FNR-010 & \textbf{\color{NavyBlue}{Satisfied}}           & \textbf{\color{NavyBlue}{Satisfied}}          & \textbf{\color{NavyBlue}{Satisfied}}          & Unknown             \\ \hline
FNR-011 & \textbf{\color{NavyBlue}{Satisfied}}          & \textbf{\color{NavyBlue}{Satisfied}}           & \textbf{\color{NavyBlue}{Satisfied}}           & \textbf{\color{NavyBlue}{Satisfied}}           \\ \hline
FNR-012 & \textbf{\color{NavyBlue}{Satisfied}}           & \textbf{\color{NavyBlue}{Satisfied}}           & Unknown             & \textbf{\color{NavyBlue}{Satisfied}}          \\ \hline
FNR-013 & \textbf{\color{Orange}{Partially}}            & \textbf{\color{NavyBlue}{Satisfied}}           & \textbf{\color{Orange}{Partially}}             & \textbf{\color{NavyBlue}{Satisfied}}              \\ \hline
FNR-014 & \textbf{\color{red}{Not Satisfied}}            & Unknown             & \textbf{\color{NavyBlue}{Satisfied}}           & \textbf{\color{Orange}{Partially}}             \\ \hline
FNR-015 & \textbf{\color{NavyBlue}{Satisfied}}           & \textbf{\color{NavyBlue}{Satisfied}}           & \textbf{\color{NavyBlue}{Satisfied}}           & \textbf{\color{NavyBlue}{Satisfied}}  \\ \hline
FNR-016 & \textbf{\color{Orange}{Partially}}           & \textbf{\color{NavyBlue}{Satisfied}}           & \textbf{\color{NavyBlue}{Satisfied}}           & Unknown             \\ \hline
\end{tabular}
}
\end{table}

From Table \ref{Table:tab1}, it can be noticed that the OM2M platform meets most of the requirements with a high degree of satisfaction. Additionally, most of the partially satisfied requirements could be satisfied by creating new plugins. This is an advantage of the modular architecture of the OM2M platform that makes it highly extensible. For example, the platform's protection of the confidentiality, integrity, availability, and privacy of the data exchanged at rest and in processing could be accomplished through creating a new plugin that would implement the required security mechanisms as part of the platform. These plugins are created and integrated in the platform to add new features that were not available by the original OM2M platform. The clear and available documentation of the standards, and the community of researchers and developers working on the OM2M platform makes such step easier. However, it is worth noting that the platform does not support streamed communication of data (i.e. video and data feeds) which renders it as a non-viable option for any application that requires streamed communication~\cite{OneM2M:Ref18}.

The IoTivity platform has a similar satisfaction level to the OM2M platform. It is based on the OCF standard that provides specifications, interoperability  guidelines and device certification program~\cite{IoTivityOCF:Ref}. Therefore, in addition to providing a platform, it could be considered a good option for device manufacturers to develop secure IoT devices that are certified by the OCF standard. In addition, the IoTivity platform is developed with the security by design approach, and the platform has its own security layer. However, it is not clear if this layer could be modified which might affect its development flexibility. Consequently, its selection as a platform for any use case could be a hard option especially if it have more strict security requirements.

The LwM2M protocol developed by OMA SpecWorks is considered as a management solution for constrained devices. The solution is not a complete standalone platform, but rather it is a protocol implemented in Anjay, Eclipse Leshan, and Eclipse Wakaama. Therefore, the protocol and these solutions do not satisfy several requirements. However, it is worth noting that the OneM2M and FIWARE standards have provided their platforms with the solution to bridge and connect a device that utilizes the LwM2M protocol~\cite{OneM2M:Ref49,FIWARE:Ref50}.

Finally, FIWARE is another platform that satisfies most of the requirements. In addition, it provides the ability to add new features through its additional complementary generic enablers. The platform provides several ready made complementary generic enablers that developers can deploy which could make the development process of the platform more efficient and easier. In other words, the developers might be able to satisfy the non-satisfied requirements. Moreover, it provides the ability to integrate the platform with third party database platforms such as Amazon AWS. Consequently, if the use case relies heavily on data processing, analysis, and visualization, the platform stands out as the best solution in this regard.

\section{Platform Selection for IoT Applications} \label{sec:usecases}
To design a horizontal IoT platform, it is required to understand the application in which the platform will be used to identify the features that the platform needs to provide, and be able to state and develop the requirements of the platform. In this work, two use cases will be used as examples to illustrate how to develop a horizontal IoT platform. The process starts by understating the use case, listing the requirements of the use case, and then a comparison between the available horizontal IoT platforms that can be used as a starting point to develop the required horizontal IoT platform for that use case. As this work is focusing on the open source horizontal IoT platforms, the considered platforms are the platforms that were introduced in Section~\ref{sec:refImp}.

\subsection{Use Case 1: E-health}
The first use case considers a horizontal IoT platform for an e-health application. In e-health applications, medical IoT devices utilize sensors and communication technologies to monitor patient's health and transfer data to a server~\cite{eldosouky2018cybersecurity}. This allows patients and physicians to track patient's health remotely. To satisfy this use case, it is very important to consider the different communication schemes and patterns that could be utilized by medical IoT devices to send their collected data securely to the horizontal IoT platform. To do that, medical devices could be categorized based on their communication strategies within the horizontal platform. Table \ref{Table:EHDevCat} summarizes the different categories for the medical IoT devices. There are two points to consider in this categorization. The first point is related to the encryption of the data exchanged between the medical IoT device and the horizontal IoT platform. For example, a device might not encrypt its own data due to its simple hardware capabilities, whereas another more powerful device might be able to encrypt its own data before it is communicated to the platform. The second point is related to the communication link between the medical IoT device and the horizontal IoT platform. The communication link with the horizontal IoT platform can be direct or indirect. In direct communication, the communication could be either a one-way communication or a two-way communication to exchange data. The indirect communication scheme considers the scenario where the medical IoT device does not communicate its data directly to the horizontal IoT platform. Instead, the device communicates its data to the manufacturer’s cloud. In this case, it is assumed that the device cannot be configured to send the data to the horizontal IoT platform directly without going through the manufacturer’s cloud. Afterwards, and through an API, data will be collected from the manufacturer’s cloud and sent to the horizontal IoT platform. Garmin wearables are examples that fit into this communication scheme~\cite{Garmin:Ref7}. These wearables collect data from users and send it to the cloud known as Garmin Connect. Afterwards, Garmin provides the developers access to this data through Garmin Health API. The API provides pre-processed JSON files that summarize individual user data uploaded to Garmin Connect from supported devices such as heart rate, stress, sleep, etc. for all-day activity.
The communication patterns for medical IoT devices fall into infrequent communications, small data transfer, and large file transfer.

\begin{table*}[t]
\centering
\caption{List of Categories for Medical IoT Devices}\label{Table:EHDevCat}
\resizebox{\textwidth}{!}{%
\begin{tabular}{c||l}
\hline
Device Category     & Category Description               \\ \hline\hline

1 & The IoT medical device communicates the data in a one-way communication directly to the horizontal IoT platform.                 \\ \hline
2 & The IoT medical device communicates encrypted data in a one-way communication directly to the horizontal IoT platform.           \\ \hline
3 & The IoT medical device communicates the data in a two-way communication directly to the horizontal IoT platform.                 \\ \hline
4 & The IoT medical device communicates encrypted data in a two-way communication directly to the horizontal IoT platform.           \\ \hline
5 & \begin{tabular}[l]{@{}l@{}}The IoT medical device communicates the data to the horizontal platform indirectly based on the use of an API. \\ The API is used by the horizontal platform to communicate with the cloud of the device manufacturer to get the data. \end{tabular}  \\ \hline
\end{tabular}
}
\end{table*}

\begin{figure*}[t]
\centering
\includegraphics[width=0.67\textwidth]{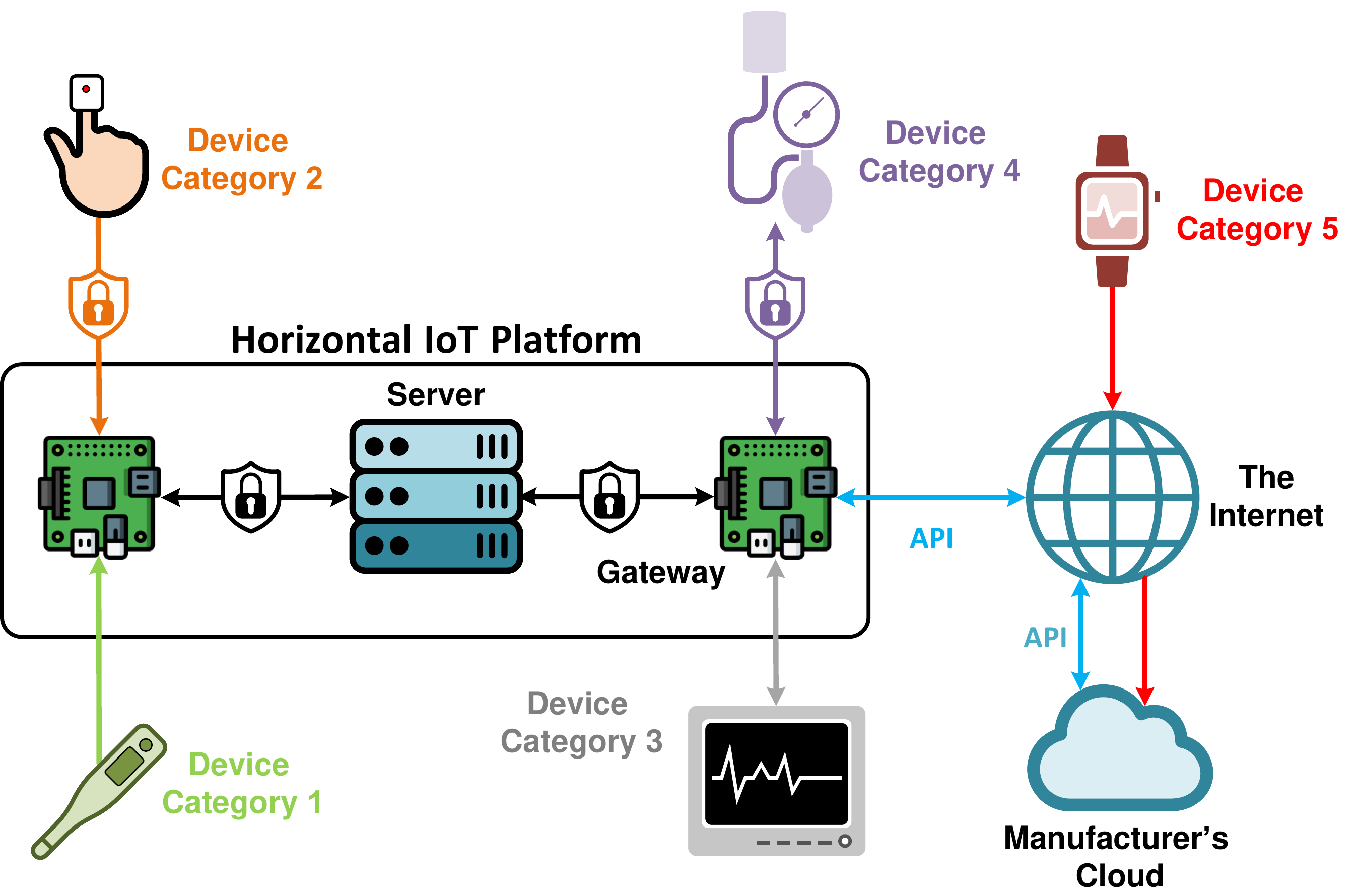}
\caption{The Horizontal IoT platform context diagram for the e-health use case.}
\label{fig:ehealthusecase}
\end{figure*}

After discussing the e-health use case, and its categories of devices, the overall platform’s functionality and major interactions within the IoT-enabled e-health platform can be presented. \figurename \ref{fig:ehealthusecase} illustrates a context diagram of the horizontal IoT platform proposed for the e-health use case. As shown in the diagram, there are four major components: Devices, Gateway, the horizontal IoT platform, and the manufacturer cloud. The devices are categorized in five different categories based on their communication strategy (see Table \ref{Table:EHDevCat}). These devices communicate with the horizontal IoT platform through a gateway. The purpose of the gateway is to collect and route data from the devices to the platform. During the routing process, these gateways could provide data formatting services. This means that they will take the data from the devices, and format it in a common way before communicating it to the horizontal IoT platform. In addition, these gateways can add another layer of data encryption. Devices in categories~1 and~3 send their data to the platform without encryption through the gateway. However, devices in categories~2 and~4 send their data to the platform after encrypting it. It is worth noting that the devices in categories~1 and~2 use one-way communication links, whereas devices in categories 3 and 4 use two-way communication links. On the other hand, devices in category 5 do not communicate with the platform directly. Therefore, the manufacturer cloud is considered in the diagram where the platform’s gateway will collect the data of the medical IoT devices using an API.

The requirements of the e-health use case is the final step before comparing the considered platform to select the most suitable platform for the development and implementation. The requirements of the e-health use case are shown in Table~\ref{Table:Reqe-health}, after categorizing the presented requirements in Section~\ref{sec:req} using MoSCoW's prioritization criteria for the functionality of the e-health application. For example, the requirements FNR-001 and FNR-004 regarding the secured seamless connection and communication of data from IoT devices to the horizontal platform is crucial to have a functioning platform. Therefore, they are considered a ``Must Have" requirements. The requirement FNR-003 that allows the platform to identify all the supported communication technologies and/or protocols by an IoT device is important with a lower degree. Hence, it is considered as a ``Should Have" requirement. On the other hand, the requirements FNR-015 and FNR-016 are not critical to the functioning of the e-health platform. The location tracking of remote devices (FNR-015) is a good feature, however, it might introduce a privacy problem to the users. Additionally, the ability to maintain devices through software updates (FNR-016) is important but it could be performed when the user is following up with their physician. Therefore, these requirements would result in a more convenient platform, however, they could be considered as ``Will Not Have" requirements.

\begin{table}[t]
\centering
\caption{The list of requirements for the e-health use case}\label{Table:Reqe-health}
\resizebox{\linewidth}{!}{%
\begin{tabular}{c||c}
\hline
Requirement Type                        & Requirement ID     \\ \hline\hline
Must Have   & \begin{tabular}[c]{@{}c@{}}FNR-001, FNR-004, FNR-005, FNR-007, FNR-008 \\ FNR-009, FNR-010, FNR-011, FNR-012\end{tabular} \\ \hline
Should Have & FNR-003, FNR-006, FNR-013  \\ \hline
Could Have    & FNR-002, FNR-014 \\ \hline
Will Not Have     & FNR-015, FNR-016 \\ \hline    
\end{tabular}
}
\end{table}

The selection of the platform to implement the e-health use case can be accomplished after identifying all aspects of the use case. Based on the comparison and discussion reported in Sections \ref{sec:comparison} and \ref{sec:discussion}, it is found that the OM2M, IoTivity and FIWARE platforms have very close satisfaction levels for the requirements, with a slight advantage to the OM2M platform.
Compared to the IoTivity platform, the OM2M platform has the same satisfaction levels of the ``Must have" requirements. However, for the ``Should Have" requirements, the OM2M platform partially satisfies the FNR-006 compared to the unknown status of IoTivity. 
Similarly, both OM2M and FIWARE have the same satisfaction levels of all the ``Must Have" requirements except for the FNR-010 where the FIWARE platform satisfaction is unknown compared to the partial satisfaction of the OM2M platform. Therefore, the OM2M platform is considered to be more suitable for the e-health use case.
In addition, the OM2M platform provides the following features:

\begin{itemize}
    \item The modular architecture of the OM2M platform that runs on the top of an OSGi layer. This makes it highly extensible and allows adding new features to the platform via plugins. This can be utilized to add new security plugins which could be developed to suit the needs of the e-health use case.

    \item It provides several basic built-in security features that can be utilized in addition to the extra security features that will be added through developed plugins.
    
    \item It provides the ability to develop services independently of the underlying network while facilitating the deployment of vertical applications and heterogeneous devices~\cite{TEF:Ref3}.
    \pagebreak
    \item It supports several protocol bindings such as HTTP, MQTT, and CoAP. Additionally, it provides various interworking proxies that enable seamless communication.
    
    \item It provides a horizontal CSE that is deployed in the server, gateway, and devices. The CSE provides application enablement, security, triggering, notification, persistency, device interworking, device management, etc.~\cite{TEF:Ref3}.

    \item It is an open source implementation of the oneM2M standards.

\end{itemize}

\subsection{Use Case 2: Smart City}
\begin{figure*}[t]
\centering
\includegraphics[width=0.7\textwidth]{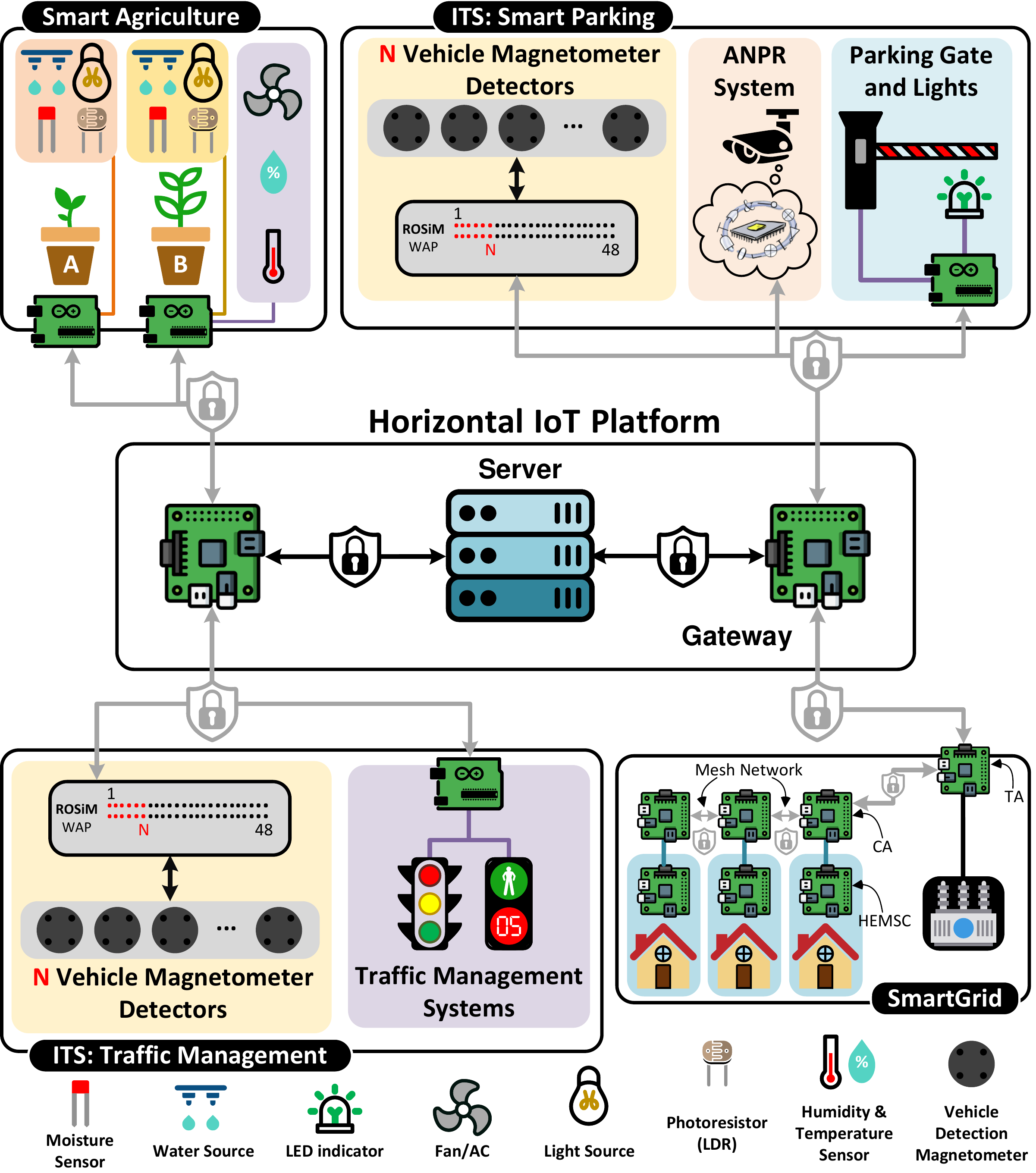}
\caption{The Horizontal IoT platform context diagram for the smart city use case.}
\label{fig:smartcityusecase}
\vspace{-0.3cm}
\end{figure*}

The second use case considers a horizontal IoT platform for a smart city application. In the smart city application, sensors and actuators are deployed around the city to monitor, control, and automate processes. This shall save resources by reducing wastes while improving the quality of services~\cite{IOT:REF51, IOT:REF54}. The smart city use case requires automation compared to the e-health use case where patients or/and physicians will read the measurements provided by the medical IoT devices to take the suitable medical action. However, in the smart city application, the system is automated where the system is controlling actuators to take actions where user interaction could be useful for some specific tasks. In this use case, the considered smart city system includes the following subsystems: smart agriculture, Intelligent Transportation Systems (ITS), and smart grid as depicted in \figurename \ref{fig:smartcityusecase}. The different components of the subsystems connect to the horizontal IoT platform through gateways. Similar to the e-health use case, these gateways could provide data formatting services, and data encryption in addition to collecting and routing data between the devices and the platform.

\begin{itemize}
    \item \textbf{Smart Agriculture}: The smart agriculture subsystem of the horizontal IoT platform utilizes different sensors to monitor the temperature, humidity, soil moisture, and lighting conditions of different plants. Then, it would compare these measurements with the optimal growing environment conditions that are pre-programmed to control the different actuators of the system to water the plant, switch on/off a light source, or cool the environment using a fan or an air conditioner. This ensures that plants are always in their optimal growing environment conditions in an automated approach. The information of this subsystem does not need to be public where the people living in the city would not benefit from this information. The communication patterns for the smart agriculture subsystem includes infrequent communications, and small data transfer.

    \item \textbf{ITS: Smart Parking}: Smart parking systems are deployed to control and manage parking areas. However, the information collected by these systems is not shared with their users which could improve the service provided by these parking areas. People around the city would be able to track real-time parking availability information around the city especially in congested areas. Different parking areas around the city might have different levels of control and monitoring (e.g., limited time parking, controlled access parking, etc.). To perform these levels of control, the following systems could be used: 
    \begin{itemize}
        \item  Smart parking systems may include Automated Number Plate Recognition (ANPR) systems that utilize a camera and system on chip circuits to identify vehicles' number plates. An example of such system could be seen in the work presented in~\cite{ANPRNPLCS,ANPROCR}.
        
        \item  Vehicle detection systems utilize vehicle magnetometer detectors to detect if a parking space is occupied. These magnetometers communicate with a Wireless Access Point (WAP) to communicate its readings. The system provided by ROSIM ITS could be considered an example of a wireless vehicle detection system that could be utilized in a smart parking application~\cite{ROSIMITS}.
        
        \item   Access control system that consists of a gate and user payment device to control the entry and exit of vehicles to the parking area.
        
    \end{itemize}
    
    In this work, the most strict case of parking control and monitoring is considered since it requires the deployment of all these systems. Therefore, the communication patterns for the smart parking subsystem includes infrequent communications, small data transfer, and large file transfer.

    \item \textbf{ITS: Traffic Management}: Traffic management is another ITS system that is deployed in smart cities to manage traffic and solve vehicular traffic congestion problems. Traffic congestion affects the quality of life and safety of a city through increasing transportation delays and fuel consumption, affecting peoples' health, and hindering emergency vehicles from providing their services. The impact of these issues could be limited or even mitigated if traffic is managed and vehicular traffic flow is maintained. The deployment of traffic management solutions allow the control of traffic signals based on the detected vehicular traffic conditions (i.e., number of vehicles, flow condition, etc.)~\cite{TRAFFICMANAGEMENT}. In this study, the considered traffic management system consists of a vehicle magnetometer detector to detect vehicular traffic condition (similar to the system used in the smart parking application), Closed-circuit Television (CCTV) system to monitor traffic, and traffic management systems to control traffic signals. The collected information from a traffic management system would help the people through managing traffic and by providing them with real-time information on the size and flow of traffic around the city. The communication patterns for the traffic management subsystem includes infrequent communications, small data transfer, large file transfer, and streamed communication.

    \item \textbf{Smart Grid}: The smart grid is a critical part of the infrastructure of smart cities. In this work, it very hard to consider and represent every part of a smart grid. Therefore, it is represented by  the system compliant with the IEEE 2030.5 smart energy profile 2.0 requirements that is presented in~\cite{SMARTGRID} and it consists of the Transformer Agent (TA), the Customer Agent (CA), and the Home Energy Management System Controller (HEMSC). It is worth noting that the HEMSC communicates with the TA through the CA. The CAs establish a mesh network where the information is transmitted through this mesh network to reach the TA. All the information communicated in the smart grid is private where only the user is able to access his/her own information. Here is a summary of each entity in the considered smart grid system:
    
    \begin{itemize}
        \item     The TA is located near the neighbourhood distribution transformer that communicates with the CA to coordinate demand response actions. In addition, it communicates with the horizontal IoT platform for commands, control and monitoring purposes. 
        
        \item     The CA is located at the house near the service meter. Its operation is to communicate with the TA to coordinate demand response actions, and interface the TA with the HEMSC to ensure secure and private communication between the utility and customer.
        
        \item     The HEMSC is located inside the house where it is responsible for all the intelligent monitoring and control of all the devices deployed in the house (e.g., appliances, generators, load controller, etc.). It hides all the details of the deployment of these devices from the utility to ensure customer privacy. In addition, it communicates with the utility after translating the information using the appropriate protocol format and command to ensure that the demand response sent by the TA to each CA is met.

    \end{itemize}
    
    \end{itemize}

    Therefore, the communication patterns for the smart grid subsystem falls into infrequent communications, and small data transfer.
    
    The requirements of the smart city use case is the final step before comparing the considered platform to select the most suitable platform for the development and implementation. The requirements of the smart city use case is shown in Table~\ref{Table:Reqsmartcity} after categorizing the presented requirements in Section~\ref{sec:req} based on their priority for the functionality of the smart city application. For example, FNR-005 and FNR-007 are considered as ``Must Have" requirements. In a smart city application, IoT devices are spread over the city where the devices are not secured and can be accessible. Therefore, it is crucial to authenticate IoT devices once they are connected to the platform and process their data to take the suitable action. On the other hand, managing traffic with different priority is less important in this application compared to the e-health application. Hence, FNR-010 is considered as a ``Should Have" requirement. Moreover, the requirements FNR-015 and FNR-016 are considered as ``Could Have" requirements in the smart city platform compared to the e-health use case. This is due to the larger number of devices deployed around the city, which makes managing these devices much harder. In addition, location tracking plays an important role in some of these application and how they are automated such as ITS.
    
\begin{table}[ht]
\centering
\caption{The list of requirements for the smart city use case}\label{Table:Reqsmartcity}
\resizebox{\linewidth}{!}{%
\begin{tabular}{c||c}
\hline
Requirement Type                        & Requirement ID     \\ \hline\hline

Must Have  & \begin{tabular}[c]{@{}l@{}}FNR-001, FNR-004, FNR-005, FNR-007, \\ FNR-008, FNR-009, FNR-011, FNR-012\end{tabular} \\ \hline
Should Have & FNR-003, FNR-006, FNR-010, FNR-014 \\ \hline
Could Have  & FNR-002, FNR-013, FNR-015, FNR-016 \\ \hline
Will Not Have  & --  \\ \hline                                          

\end{tabular}
}
\end{table}

    The selection of the platform to implement the smart city use case can be accomplished after identifying all aspects of the use case. Based on the comparison and discussion reported in Sections~\ref{sec:comparison}~and~\ref{sec:discussion}, it seems that the OM2M platform is the most suitable option. However, the limitation where the OM2M platform does not support streamed communication pattern renders it unsuitable for this use case. In addition, the OM2M platform is lacking ready made plugins to add features to the platform. Therefore, the FIWARE platform is found to be the most suitable option due to the following features: 
    
\begin{itemize}
    \item The architecture of the FIWARE platform that allows the developers to add the required features to the platform through new or ready made generic enablers that integrate with the FIWARE context broker (the only mandatory component of the platform).
    
    \item The ability to interface the platform with third party systems and databases to process, analyze and visualize collected data to take smart decision and meet the expected smart behavior.

    \item The IoT agents provided by IDAS generic enabler that interface IoT devices utilizing the most widely used protocols such as HTTP, MQTT, and LwM2M with the platform to enable seamless communication.
    
    \item It is an open source implementation of the FIWARE standards.

\end{itemize}

\begin{table*}[t]
\caption{The list of acronyms}\label{Table:acronym}
\resizebox{\textwidth}{!}{%
\begin{tabular}{l|l||l|l}
\hline
Acronym & Meaning                                   & Acronym   & Meaning                                                                                              \\ \hline\hline
ABAC    & Attribute-Based Access Control           & JSON      & JavaScript Object Notation                                                                           \\
ACE     & Access Control Element                   & LAAS-CNRS & Laboratory for Analysis and Architecture of Systems - Centre   National de la Recherche Scientifique \\
ACL     & Access Control List                      & LwM2M     & Lightweight M2M                                                                                      \\
ACP     & Access Control Polices                   & M2M       & Machine-to-Machine                                                                                   \\
ANPR    & Automated Number Plate Recognition       & MIC       & Message Integrity Code                                                                               \\
APIs    & Application Programming Interfaces       & MQTT      & Message Queuing Telemetry Transport                                                                  \\
ASE     & Asymmetric Secure Element                & OAuth2    & Open Authorization 2                                                                                 \\
CA      & Customer Agent                           & OCF       & Open Connectivity Foundation                                                                         \\
CBOR    & Concise Binary Object Representation     & OSGi      & Open Service Gateway Initiative                                                                      \\
CCTV    & Closed-circuit Television                & OTGC      & Onboarding Tool and Generic Client                                                                   \\
CoAP    & Constrained Application Protocol         & RBAC      & Role-Based Access Control                                                                            \\
CSE     & Common Services Entity                   & SBAC      & Subject-Based Access Control                                                                         \\
DTLS    & Datagram Transport Layer Security        & SSID      & Service Set IDentifier                                                                               \\
HEMSC   & Home Energy Management System Controller & TA        & Transformer Agent                                                                                    \\
HTTP    & HyperText Transfer Protocol              & TLS       & Transport Layer Security                                                                             \\
ID      & Identification number                    & TLS-PSK   & Transport Layer Security Pre-Shared Key cipher suites                                                \\
IoT     & Internet of Things                       & UML       & Unified Modeling Language                                                                            \\
IP      & Internet Protocol                        & WAP       & Wireless Access Point                                                                                \\
ITS     & Intelligent Transportation Systems       & XACML     & eXtensible Access Control Markup Language                                        
        \\ \hline
\end{tabular}
}
\end{table*}

\section{Conclusions} \label{sec:conclusion}
IoT platforms plays an important role in the deployment of IoT solutions in different applications. Therefore, this paper introduced a guideline for developers and researchers to address the challenge of selecting the most suitable open source horizontal IoT platform as a starting point for their development. The guideline starts by identifying a list of functional requirements that are essential for most IoT applications. The provided list of requirements may be modified by the developers to suite their application accordingly. Afterwards, the considered open source horizontal IoT platforms are surveyed and evaluated based on the identified requirements. The aim of this comparison is to recognize the capabilities and limitations of each platform to select the most suitable option. This option is considered as a starting point, which requires additional contribution from the developers and researchers to ensure that the proposed IoT system satisfies the considered use case. In addition, the e-health and smart city use cases were introduced as examples on how to consider the proposed guideline to identify the most suitable horizontal IoT platform as a starting point.

\begin{appendix}
The acronyms used in this article are listed in Table \ref{Table:acronym}.
\end{appendix}

\bibliography{references.bib}
\bibliographystyle{IEEEtran}
\end{document}